\documentclass[10pt]{article}
\usepackage{graphicx}

\usepackage{cite}

\usepackage{times}
\newenvironment{sciabstract}{%
\begin{quote} \bf}
{\end{quote}}

\newcounter{lastnote}
\newenvironment{scilastnote}{%
\setcounter{lastnote}{37}%
\begin{list}%
{\arabic{lastnote}.}
{\setlength{\leftmargin}{.22in}}
{\setlength{\labelsep}{.5em}}}
{\end{list}}

\title{The First Stars in the Universe and Cosmic Reionization}

\author
{Rennan Barkana$^{1}$\\
\\
\normalsize{$^{1}$School of Physics and Astronomy,}\\ 
\normalsize{The Raymond and Beverly Sackler Faculty of Exact Sciences,}\\ 
\normalsize{Tel Aviv University, Tel Aviv 69978, ISRAEL}\\
\normalsize{E-mail: barkana@wise.tau.ac.il.}
}

\date{}

\begin{document}




\maketitle


\begin{sciabstract}
The earliest generation of stars, far from being a mere novelty,
transformed the universe from darkness to light. The first atoms to
form after the Big Bang filled the universe with atomic hydrogen and a
few light elements. As gravity pulled gas clouds together, the first
stars ignited and their radiation turned the surrounding atoms into
ions. By looking at gas between us and distant galaxies, we know that
this ionization eventually pervaded all space, so that few hydrogen
atoms remain today between galaxies. Knowing exactly when and how it
did so is a primary goal of cosmologists, because this would tell us
when the early stars formed and in what kinds of galaxies. Although
this ionization is beginning to be understood by using theoretical
models and computer simulations, a new generation of telescopes is
being built that will map atomic hydrogen throughout the universe.
\end{sciabstract}

Astronomers engage in cosmic archaeology. The farther away they look
in distance, the further back in time they see, because of the time it
takes light to travel from distant stars and galaxies to Earth
today. In principle, this allows the entire 13.7 billion year cosmic
history of our universe to be reconstructed by surveying the galaxies
and other sources of light to large distances (Fig.~1). To measure
distance, astronomers use the characteristic emission patterns of
hydrogen and other chemical elements in the spectrum of each galaxy to
measure its cosmological redshift $z$. As the universe expands, light
wavelengths get stretched as well, so that the spectrum we observe
today is shifted from the emitted one by a factor of $1+z$ in
wavelength. This then implies that the universe has expanded by a
factor of $1+z$ in linear dimension since that time, and cosmologists
can calculate the corresponding distance and cosmic age. Large
telescopes have allowed astronomers to observe faint galaxies that are
so far away that we see them more than ten billion years in the
past. So galaxies were in existence as early as 850 million years
after the Big Bang, at a redshift of $z \sim 6.5$ (1, 2).

\begin{figure}[]
\centering
\includegraphics[width=5in]{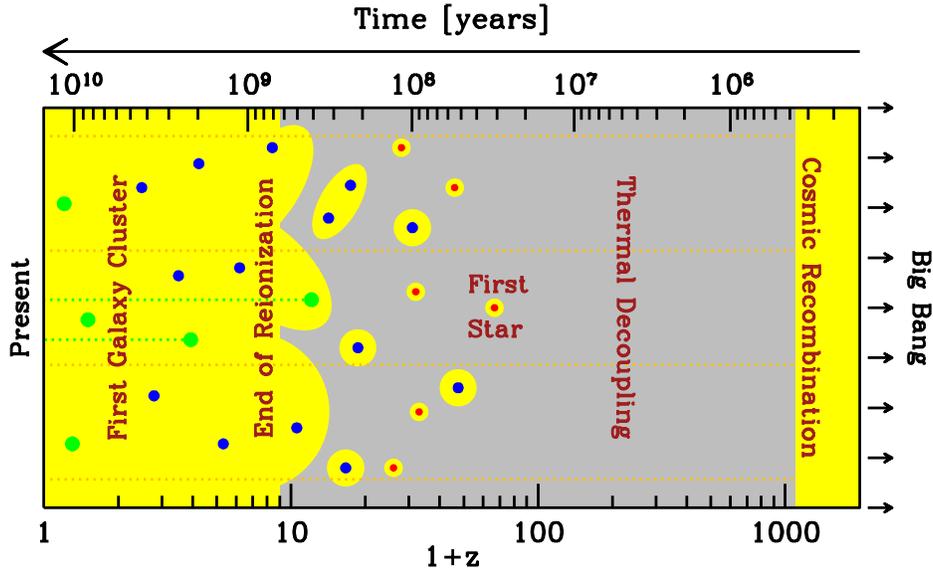}
\caption{\footnotesize Overview of cosmic history, with the age of the
universe shown on the top axis and the corresponding redshift on the
bottom axis. Yellow represents regions where the hydrogen is ionized,
and gray, neutral regions. Stars form in galaxies located within dark
matter concentrations whose typical size grows with time, starting
with 300 light years (red circles) for the host of the first star,
rising to 3000 light years (blue circles) for the sources of
reionization, and reaching 600000 light years (green circles) for
present-day galaxies like our own Milky Way. Astronomers probe the
evolution of the cosmic gas using the absorption of background light
(dotted lines) by atomic hydrogen along the line of sight. The
classical technique uses absorption by the Ly resonance of hydrogen of
the light from bright quasars located within massive galaxies, while a
new type of astronomical observation will use the 21-cm line of
hydrogen with the CMB as the background source.}
\end{figure}

In addition to galaxies, the other major probe of observational
cosmology is the cosmic microwave background (CMB), a radiation relic
from the fiery beginning of the universe. The universe cools as it
expands, so it was initially far denser and hotter than it is
today. For hundreds of thousands of years the cosmic gas consisted of
a plasma of protons, electrons, and light nuclei, sustained by the
intense thermal motion of these particles. Just like the plasma in our
own Sun, the ancient cosmic plasma emitted and scattered a strong
field of visible and ultraviolet photons. About 400,000 years after
the Big Bang, however, the temperature of the universe dipped for the
first time below a few thousand degrees Kelvin. The protons and
electrons were now moving slowly enough that they could attract each
other and form hydrogen atoms, in a process known as cosmic
recombination. With the scattering of the energetic photons now much
reduced, the photons continued traveling in straight lines, mostly
undisturbed except that cosmic expansion has redshifted them into the
microwave regime. The emission temperature of the observed spectrum of
these CMB photons is the same in all directions to one part in 100000,
which reveals that conditions were extremely uniform in the early
universe.

It was at the moment of cosmic recombination that gravity entered the
scene. Since that time, gravity has amplified the tiny fluctuations in
temperature and density observed in the CMB data (3). Regions that
started out slightly denser than average began to contract because the
gravitational forces were also slightly stronger than average in these
regions. Eventually, after millions of years of contraction, galaxies
and the stars within them were able to form. This process, however,
would have taken too long to explain the abundance of galaxies today,
if it involved only the observed cosmic gas. Instead, gravity is
strongly enhanced by the presence of dark matter -- an unknown
substance that makes up the vast majority (85\%) of the cosmic density
of matter. The motion of stars and gas around the centers of nearby
galaxies indicates that each is surrounded by an extended mass of dark
matter, and so dark matter concentrations are generally referred to as
halos.

Gravity explains how some gas is pulled into the deep potential wells
within dark matter halos and forms the galaxies. On the other hand,
cosmologists at first expected the gas outside halos to remain mostly
undisturbed. However, observations show that it has not remained
neutral (i.e., in atomic form) until the present. To learn about
diffuse gas pervading the space outside and between galaxies (referred
to as the intergalactic medium, or IGM), astronomers study its
absorption against distant quasars, the brightest known astronomical
objects. Quasars' great luminosities are believed to be powered by
black holes weighing a billion times the mass of the Sun that are
situated in the dense centers of massive galaxies. As the surrounding
gas falls in toward the black hole, violent collisions within the gas
blast radiation into space, creating a beacon visible from afar.

The Lyman-alpha (Ly$\alpha$) resonance line of hydrogen at a
wavelength of 1216 \AA\ has been widely used to trace hydrogen gas
through its absorption of quasar light (4). The expansion of the
universe gives this tool an important advantage common to all spectral
absorption probes. Since the wavelength of every photon grows as the
universe expands, the rest-frame absorption at 1216 \AA\ by a gas
element at redshift $z$ is observed today at a wavelength of 1216$\,
(1+z)$ \AA. The absorptions of the different gas elements along the
line of sight are therefore distributed over a broad range of
wavelengths, making it possible to measure the distribution of
intergalactic hydrogen.

Ly$\alpha$ absorption shows that the IGM has been a hot plasma at
least from a cosmic age of 850 million years ($z \sim 6.5$) until
today (2). Thus, the hydrogen must have been ionized for a second time
after it became neutral at cosmic recombination. Radiation from the
first generations of stars is a plausible source for the ionizing
photons that transformed the IGM by reionizing the hydrogen throughout
the universe.

Absorption at Ly$\alpha$ is so efficient that it becomes difficult to
use as observations approach the reionization epoch where the density
of neutral hydrogen is higher than at more recent times (2). As
described below, cosmologists believe that the new technique of
``21-cm cosmology'' will allow them to measure how the reionization
process developed over time and to test theoretical predictions of the
properties of the first stars (5).

\section*{The First Stars}

The most distant galaxies that we can see today were already shining
brightly when the universe was just a billion years old. In the
``hierarchical model'' of galaxy formation - in which small galaxies
form first and then merge or accrete gas to form larger galaxies -
smaller objects should have formed even earlier. So what were the
smallest units within which the first stars formed?

Stars form when huge amounts of matter collapse to enormous
densities. However, the process can be stopped if the pressure exerted
by the hot intergalactic gas prevents outlying gas from falling into
dark matter concentrations. As the gas falls into a dark matter halo,
it forms shocks due to unstable supersonic flow and in the process
heats up and can only collapse further by first radiating its energy
away. This restricts this process of collapse to very large clumps of
dark matter that are around 100,000 times the mass of the Sun. Inside
these clumps, the shocked gas loses energy by emitting radiation from
excited molecular hydrogen that formed naturally within the primordial
gas mixture of hydrogen and helium (6-7).

Advances in computing power have made possible detailed numerical
simulations of how the first stars formed (8-9). These simulations
begin in the early universe, in which dark matter and gas are
distributed uniformly, apart from tiny variations in density and
temperature that are statistically distributed according to the
patterns observed in the CMB. In order to span the vast range of
scales needed to simulate an individual star within a cosmological
context, the newest code follows a box a million light years in length
and zooms in repeatedly on the densest part of the first collapsing
cloud that is found within the simulated volume. The simulation
follows gravity, hydrodynamics, and chemical processes in the
primordial gas, and resolves a scale 10 orders of magnitudes smaller
than that of the simulated box. While the resolved scale is still
three orders of magnitudes larger than the size of the Sun, the
simulations indicate that the first stars most likely weighed $\sim
100 M_{\odot}$ each.

To estimate when the first stars formed we must remember that the
first 100,000 solar mass halos collapsed in regions that happened to
have a particularly high density enhancement very early on. There were
initially only a few such regions in the entire universe, so a
simulation that is limited to a small volume is unlikely to find such
halos until much later. Simulating the entire universe is well beyond
the capabilities of current simulations, but analytical models predict
that the first observable star in the universe probably formed 30
million years after the Big Bang (10), less than a quarter of one
percent of the universe's total age of 13.7 billion years.

Although stars were extremely rare at first, gravitational collapse
increased the abundance of galactic halos and star formation sites
with time (Fig.~1). The sources of reionization were most likely a
second generation of larger galaxies that formed in halos of mass
above $\sim 10^7 M_{\odot}$ (11). The first Milky-Way-sized halo $M =
10^{12} M_{\odot}$ is predicted to have formed 400 million years after
the Big Bang (10), but such halos have become typical galactic hosts
only in the last five billion years.

\section*{Cosmic Reionization}

Given the understanding described above of how many galaxies formed at
various times, the course of reionization can be determined
universe-wide by counting photons from all sources of light
(12-17). Both stars and black holes contribute ionizing photons, but
the early universe is dominated by small galaxies which in the local
universe have central black holes that are disproportionately
small. Thus, stars most likely dominated the production of ionizing
photons during the reionization epoch. If the stars within the early
galaxies were similar to those observed today, then each star produced
$\sim$4000 ionizing photons per baryon, which means that it was
sufficient to accumulate a small fraction (of order 0.1\%) of the
total gas mass in the universe into galaxies in order to ionize the
entire IGM. Since most stellar ionizing photons are only slightly more
energetic than the 13.6 eV ionization threshold of hydrogen, they are
absorbed very quickly once they reach a region with substantial
neutral hydrogen. This makes the IGM during reionization a two-phase
medium characterized by highly ionized regions separated from neutral
regions by sharp ionization fronts (see Fig.~2).

From absorption line work we see that the IGM is completely ionized
towards even the most distant objects (quasars) known. There are
hints, however, that some large neutral H regions persist at these
times (18-19) so this suggests that we may not need to go to much
higher redshifts to begin to see the last stages of reionization. We
know that the universe could not have fully reionized earlier than an
age of 300 million years, since the freshly created plasma at
reionization re-scattered some of the CMB photons and created a
clearly-detected polarization signature (3) that constrains the
reionization redshift; an earlier reionization, when the universe was
still relatively dense, would have created stronger scattering than
observed.

A detailed picture of reionization as it happens will teach us a great
deal, since the spatial distribution of ionized bubbles is determined
by clustered groups of galaxies and not by individual galaxies. At
such early times galaxies were strongly clustered even on very large
scales (10 to 100 million light years), and these scales therefore
dominate the structure of reionization (20). The basic idea is simple
(21). At high redshift, galactic halos are rare and correspond to
rare, high density peaks. As an analogy, imagine searching on Earth
for mountain peaks above 5000 meters. The 200 such peaks are not at
all distributed uniformly but instead are found in a few distinct
clusters on top of large mountain ranges. Similarly, in order to find
the early galaxies, one must first locate a region with a large-scale
density enhancement, and then galaxies will be found there in
abundance.

The ionizing radiation emitted from the stars in each galaxy initially
produces an isolated ionized bubble. However, in a region dense with
galaxies the bubbles quickly overlap into one large bubble, completing
reionization in this region while the rest of the universe is still
mostly neutral (Fig.~2). Most importantly, since the abundance of rare
density peaks is very sensitive to small changes in the density
threshold, even a large-scale region with a small enhanced density
(say, 10\% above the mean density of the universe) can have a much
larger concentration of galaxies than in other regions (e.g., a 50\%
enhancement). On the other hand, reionization is harder to achieve in
dense regions, since the protons and electrons collide and re-form
hydrogen atoms more often in such regions, and newly-formed atoms need
to be reionized again by additional ionizing photons. However, the
overdense regions end up reionizing first since the number of ionizing
sources in these regions is increased so strongly (20). This is a key
prediction awaiting observational testing.

\begin{figure}[]
\centering
\includegraphics[width=4in]{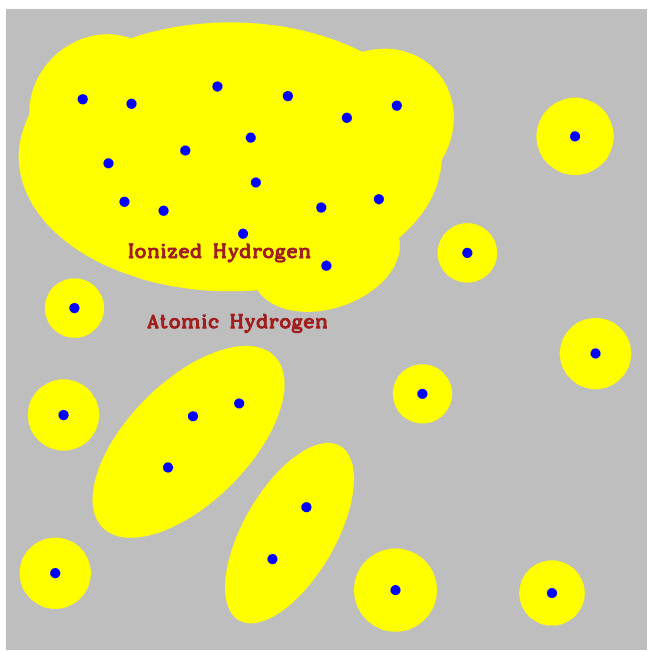}
\hfill
\includegraphics[width=3.8in]{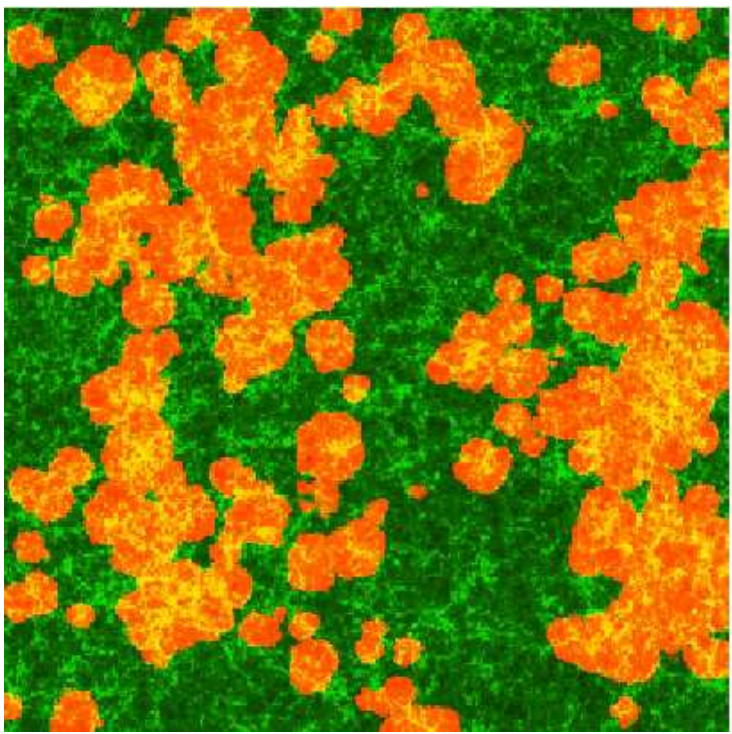}
\caption{\footnotesize The spatial structure of cosmic
reionization. The illustration (left panel) shows how regions with
large-scale overdensities form large concentrations of galaxies (dots)
whose ionizing photons produce enormous joint ionized bubbles (upper
left). At the same time, galaxies are rare within large-scale voids,
in which the IGM is still mostly neutral (lower right). A numerical
simulation of reionization (right panel, from (24)) displays a similar
variation in the sizes of ionized bubbles (orange), shown overlayed on
the density distribution (green).}
\end{figure}

Detailed analytical models that account for large-scale variations in
the abundance of galaxies (22) confirm that the typical bubble size
starts below a million light years early in reionization, as expected
for an individual galaxy, rises to 20 to 30 million light years during
the central phase (i.e., when the universe is half ionized), and then
by another factor of $\sim$5 towards the end of reionization. These
scales are given in ``comoving'' units that scale with the expansion
of the universe, so that the actual sizes at a redshift $z$ were
smaller than these numbers by a factor of $1+z$. Numerical simulations
have only recently begun to reach the enormous scales needed to
capture this evolution (23-25). Accounting precisely for gravitational
evolution on a wide range of scales but still crudely for gas
dynamics, star formation, and the radiative transfer of ionizing
photons, the simulations confirm that the large-scale topology of
reionization can be used to study the abundance and clustering of the
ionizing sources (Figs.~2 and 3).

\begin{figure}[]
\centering
\includegraphics[width=6.5in]{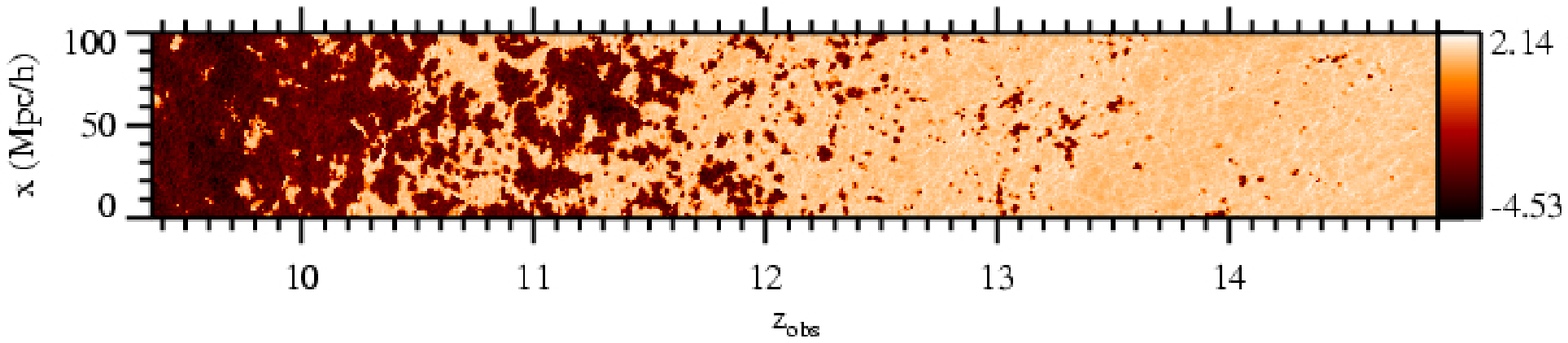}
\caption{\footnotesize Close-up of cosmic evolution during the epoch
of reionization, as revealed in a predicted 21-cm map of the IGM based
on a numerical simulation, from (24). This map is constructed from
slices of the simulated cubic box of side 450 million light years (in
comoving units), taken at various times during reionization, which for
the parameters of this particular simulation spans a period of 250
million years from redshift 15 down to 9.3. The vertical axis shows
position $\chi$ in units of Mpc/h, which equals $\sim$4.5 million
light years. This two-dimensional slice of the sky (one linear
direction on the sky versus the line-of-sight or redshift direction)
shows $\log_{10}(T_b)$, where $T_b$ (in mK) is the 21-cm brightness
temperature relative to the CMB. Since neutral regions correspond to
strong emission (i.e., a high $T_b$), this slice illustrates the
global progress of reionization and the substantial large-scale
spatial fluctuations in reionization history. Observationally it
corresponds to a narrow strip half a degree in length on the sky
observed with radio telescopes over a wavelength range of 2.2 to 3.4 m
(with each wavelength corresponding to 21-cm emission at a specific
line-of-sight distance and redshift).}
\end{figure}

\section*{21-cm Cosmology}

The prospect of studying reionization by mapping the distribution of
atomic hydrogen across the universe using its prominent 21-cm spectral
line (26) has motivated several teams to design and construct arrays
of low-frequency radio telescopes; the Primeval Structure Telescope,
the Mileura Widefield Array, and the Low Frequency Array will search
over the next few years for 21-cm emission or absorption from $z\sim
6.5$ and above, redshifted and observed today at relatively low
frequencies which correspond to wavelengths of 1.5 to 4 meters.

The idea is to use the resonance associated with the hyperfine
splitting in the ground state of hydrogen. The state with parallel
spins of the proton and electron has a slightly higher energy than the
state with anti-parallel spins, yielding a spin-flip transition which
corresponds to 21-cm wavelength radiation. While the CMB spectrum
peaks at a wavelength of 2 mm, it provides a still-measurable
intensity at meter wavelengths that can be used as the bright
background source against which we can see the expected 1\% absorption
by neutral hydrogen along the line of sight (27-28). Since the CMB
covers the entire sky, a complete three-dimensional map of neutral
hydrogen can in principle be made from the sky position of each
absorbing gas cloud together with its redshift $z$.

Because the smallest angular size resolvable by a telescope is
proportional to the observed wavelength, radio astronomy at
wavelengths as large as a meter has remained relatively
undeveloped. Producing resolved images even of large sources such as
cosmological ionized bubbles requires telescopes kilometers in
diameter. It is much more cost-effective to use a large array of
thousands of simple antennas distributed over several kilometers. The
new experiments are being placed mostly in remote sites, because the
cosmic wavelength region overlaps with more mundane terrestrial
telecommunications.

Since the 21-cm line intensity is observed with the CMB as the bright
source, the hydrogen gas produces absorption if it is colder than the
CMB and excess emission if it is hotter. The observed intensity
$I_{\nu}$ relative to the CMB at a frequency $\nu$ is measured by
radio astronomers as an effective ``brightness temperature'' $T_b$ of
blackbody emission at this frequency, defined using the Rayleigh-Jeans
limit of the Planck radiation formula: $I_{\nu} \equiv 2 k_B T_b \nu^2
/ c^2 $.

In approaching redshifted 21-cm observations, although the first
inkling might be to consider the mean emission signal, the signal is
100,000 times weaker than foreground emission from magnetized plasma
in our own Milky Way and other nearby galaxies. Thus cosmologists have
focused on the expected characteristic variations in $T_b$, both with
position on the sky and especially with frequency, which signifies
redshift for the cosmic signal but simply frequency for the
foreground, which is expected to have the known smooth spectrum of
emission from energetic particles in plasmas (29). The 21-cm
brightness temperature depends on the density of neutral hydrogen. As
explained before, large-scale patterns in the reionization are driven
by spatial variations in the abundance of galaxies; the 21-cm
fluctuations reach $\sim$5 mK (root mean square) in brightness
temperature (Fig.~3) on a scale of 30 million light years
(comoving). While detailed maps will be difficult to extract due to
the foreground emission, a statistical detection of these fluctuations
(through the power spectrum) is expected to be well within the
capabilities of the first-generation experiments now being built
(30-31). The key information on the topology and timing of
reionization can be extracted statistically, with an important check
made possible by measuring the particular form of anisotropy, expected
in the 21-cm fluctuations, that is caused by gas motions along the
line of sight (32-34).

The theoretical expectations presented here for reionization and for
the 21-cm signal are based on rather large extrapolations from
observed galaxies to deduce the properties of much smaller galaxies
that formed at an earlier cosmic epoch. Considerable surprises are
thus possible, such as an early population of quasars or even exotic
particles that emitted ionizing radiation when they radioactively
decayed. In any case, a detection of the cosmological 21-cm signal
will open a new window on the universe and likely motivate a second
generation of more powerful telescopes. These will be used to obtain
three-dimensional maps of atomic hydrogen during reionization as well
as statistical power-spectrum measurements at even higher redshifts
(using wavelengths at which the foregrounds are brighter and thus more
difficult to remove). Since the 21-cm measurements are sensitive to
any difference between the hydrogen temperature and the CMB
temperature, the potential reach of 21-cm cosmology extends down to a
cosmic age of $\sim$6 million years, when the IGM first cooled below
the CMB temperature (an event referred to as ``thermal decoupling'')
due to the cosmic expansion. The 21-cm technique may open up entirely
new cosmological areas, such as investigations of the primordial
origin of spatial density variations by measuring them on very small
scales which are washed out in the CMB (35), and the detection and
study of the galaxies from early times (200 million years) through the
effect of their stellar radiation on the intergalactic hydrogen
(36). Understandably, astronomers are eager to start tuning into the
cosmic radio channels of 21-cm cosmology.

\bibliography{scibib}

\bibliographystyle{Science}

\newpage

\end{document}